# Correcting on-chip distortion of control pulses with silicon spin qubits

Ming Ni,[1,2] Rong-Long Ma,[1,2] Zhen-Zhen Kong,[3] Ning Chu,[1,2] Wei-Zhu Liao,[1,2] Sheng-Kai Zhu,[1,2] Chu Wang,[1,2] Gang Luo,[1,2] Di Liu,[1,2] Gang Cao,[1,2,4] Gui-Lei Wang,[3,4,5,*] Hai-Ou Li,[1,2,4,*] and Guo-Ping Guo[1,2,4,6]

[1] *CAS Key Laboratory of Quantum Information, University of Science and Technology of China, Hefei, Anhui 230026, China*

[2] *CAS Center for Excellence in Quantum Information and Quantum Physics, University of Science and Technology of China, Hefei, Anhui 230026, China*

[3] *Integrated Circuit Advanced Process R&D Center, Institute of Microelectronics, Chinese Academy of Sciences, Beijing 100029, P. R. China*

[4] *Hefei National Laboratory, University of Science and Technology of China, Hefei 230088, China*

[5] *Beijing Superstring Academy of Memory Technology, Beijing 100176, China*

[6] *Origin Quantum Computing Company Limited, Hefei, Anhui 230026, China*

[*]Corresponding authors: guilei.wang@bjsamt.org.cn; haiouli@ustc.edu.cn;

Pulse distortion, as one of the coherent error sources, hinders the characterization and control of qubits. In the semiconductor quantum dot system, the distortions on measurement pulses and control pulses disturb the experimental results, while no effective calibration procedure has yet been reported. Here, we demonstrate two different calibration methods to calibrate and correct the distortion using the two-qubit system as a detector. The two calibration methods have different correction accuracy and complexity. One is the coarse predistortion (CPD) method, with which the distortion is partly relieved. The other method is the all predistortion (APD) method, with which we measure the transfer function and significantly improve the exchange oscillation homogeneity. The two methods use the exchange oscillation homogeneity as the metric and are appropriate for any qubit that oscillates with a diabatic pulse. With the APD procedure, an arbitrary control waveform can be accurately delivered to the device, which is essential for characterizing qubits and improving gate fidelity.

## I. INTRODUCTION

Universal quantum computation requires single-qubit gates and two-qubit gates with long coherence times and high gate fidelities [1-4]. When applying all these gate operations, accurately delivering the waveform to the device is necessary for the qubits defined in semiconductor quantum dots [5-7]. Optimizing the control waveform as an effective method to improve gate fidelity widely exists while realizing qubit gates [8-12]. However, the distortion of the designed waveform is inevitable in the measurement system, which, on the one hand, hinders the realization of high-fidelity qubit gates and, on the other hand, impacts the characterization of the coherence time and influences further research on the error source [13-16].

Pulse distortion in the measurement system can be induced by the narrow Arbitrary Waveform Generator (AWG) bandwidth, high-pass filtering, low-pass filtering, skin effect, impedance mismatch, etc. Generally, pulse distortion can be described by a transfer function $h(t)$, which describes the influences of all distortion sources in the control line, from room temperature electronics to a qubit on a chip. Therefore, once the transfer function is calibrated, we can apply predistortion corrections on the control pulse and precisely compensate for the distortion. It is easy to measure the pulse distortion at room temperature with a vector network analyzer or an oscilloscope, while at the cryogenic temperature, the frequently used method is to directly detect the response of the control pulse using the qubit as the sensor. For the superconductor qubit, various processes for calibrating the waveform distortion have been performed [17-23]. However, for a single spin qubit in semiconductor quantum dots, a compatible method is still lacking.

The two-qubit exchange oscillation driven by a diabatic pulse is particularly sensitive to pulse distortion. In this article, we characterize the on-chip distortion of a control pulse with the exchange oscillation of a two-qubit system as the detector. First, we introduce a CPD method and calibrate the distortion mainly induced by the bias-tee. After that, we give an APD method with a higher correction accuracy and higher complexity. The accuracy of this procedure is predominantly limited by the coherence time and oscillation frequency of the exchange oscillation. With the corrected waveform, we observed a significant improvement in the homogeneity of exchange oscillation, which indicates the effectiveness of these procedures.

## II. RESULTS AND DISCUSSION

### A. Experimental setup

We perform our experiment on a gate-defined two-qubit device fabricated on a 70 nm isotopically purified $^{28}$Si epilayer [24-26]. As shown in Fig. 1(a), the left single spin qubit $Q_L$ and the right single spin qubit $Q_R$ are located underneath the plunger gates PL and PR, respectively. The electrons tunnel in and out between the quantum dots and the electron bath underneath the reservoir RG. The electron tunneling rate can be controlled by the voltage on gate BL. The DC voltages are exerted on the plunger gates to form the potential well, which trap the electrons. During the experiments, the detuning $\varepsilon$ between the two qubits can be controlled by modifying the relative gate voltages on the plunger gates PL and PR, and the tunnel coupling $t_c$ between qubits can be tuned by the gate voltage on BR.

In Fig. 1(a), we also illustrate the measurement setup. In our system, while controlling the spin qubit, a waveform on the order of kHz is necessary to realize single-shot readout and initialization. Meanwhile, a microwave and a pulse on the order of GHz are also required to demonstrate single-qubit gates and two-qubit gates. To generate all these control signals, we use two different AWGs, AWG1 and AWG2, which generate pulses on the order of kHz and GHz, respectively. And two vector microwave sources MW1 and MW2 are applied to drive the qubits $Q_L$ and $Q_R$, respectively. As shown in Fig. 1(a), to simultaneously deliver the pulses and microwaves to the device, two bias tees that connect to the plunger gates are placed at the cryogenic temperature to combine these control signals. The pulse on the kHz order overlaps with the DC voltage at room temperature and is delivered to the device through the DC port of the bias tee. And the pulse on the GHz order overlaps with the microwaves at room temperature and is delivered to the device through the RF port of the bias tee. Among all the input signals, the distortion of the GHz pulses, which manifests as errors, are applied through the RF port of the bias tee to the qubits and impact the control process. In our method, we focus on the distortion of the control pulse on the GHz order, which is applied to drive the two-qubit gate.

### B. Exchange oscillation

For the two qubits $Q_L$ and $Q_R$ formed in two adjacent quantum dots, their Zeeman splitting energies $E_z$ are different due to the Stark shift [5-7, 27]. Meanwhile, the exchange coupling $J$ always exists and can be tuned by the detuning $\varepsilon$. Here, we focus on the basis of antiparallel states $|\uparrow\downarrow\rangle$ and $|\downarrow\uparrow\rangle$, and the Hamiltonian [28] is given as

$$H = \Delta E_z \hat{\sigma}_z + J \hat{\sigma}_x, \tag{1}$$

where $\Delta E_z$ is the Zeeman splitting energy difference for the two qubits. As we mentioned before, the detuning $\varepsilon$ between two qubits can be tuned by the relative gate voltage between $V_{\text{PL}}$ and $V_{\text{PR}}$. When the diabatic pulses are exerted on the plunger gates, due to the modified $\varepsilon$, the two-qubit state stays still while the eigenstates change, as shown in Fig. 1(b) [29]. The exchange coupling between qubits is driven with the frequency $f_{ex} = \sqrt{\Delta E_z^2 + J^2}$. For the reason of $J \propto \exp(\varepsilon)$, as long as the top edge of the diabatic pulse is flat, we should observe a homogeneous exchange oscillation with $f_{ex}$. However, as illustrated in Fig. 1(c), a slight distortion of the top edge will change the pulse amplitude as well as $\varepsilon$ at the operation point, such that $f_{ex}$ is changed and results in an apparent inhomogeneous oscillation. Regarding the exchange oscillations in our experiment, when the top edge of the fast pulse is written flat, we observe that the exchange oscillation frequency gradually slows down, as shown in the un-predistorted (UPD) situation in Fig. 4(a). This indicates that the control pulse is distorted and that the top edge decays with time. We ascribe the distortion to the RC filter circuit inside the bias tee, which will lead to the top edge of a square pulse decay.

### C. CPD method

The pulse disturbed by the RC filter circuit is illustrated in Fig. 2(a). For an input pulse $U_{in}(t)$, for example, a diabatic square input pulse in Fig. 2(a), the output pulse $U_{out}(t)$ without the distortion should rise first and then stay at a stable amplitude $A_{in}$, which is the amplitude of $U_{in}(t)$. However, distorted by the RC filter circuit, the response output pulse is different. After the rising edge, the top edge of the output pulse $U_{out}(t)$ will decay exponentially with the form $U_{out}(t) = A_{in}e^{-\frac{t}{\tau}}$, where $t$ is the duration time of the top edge and $\tau$ is the time constant, which depends on the capacitor and resistor in the RC circuit. Subsequently, after the falling edge, $U_{out}(t)$ will fall immediately and then exponentially rise to zero. In our experiment, to predistort $U_{in}(t)$, we use a pulse with a linearly ramped top edge $U_{PD,in}(t) = A_{in}(1 + kt)$. As illustrated in Fig. 2(a), the linearly ramped top edge can effectively compensate for the decay induced by the RC filter circuit. The complete correction of the bottom edge requires the predistortion pulse to hold on to a nonzero voltage level $A_{in}kt$ after the falling edge. Considering the inevitable zero voltage level after the falling edge, we linearly decrease the voltage to the zero voltage level and ignore the effect of the remaining distortion.

To detect the ramp rate $k$, we measure the exchange oscillation as $k$ increases. The time domain of the oscillation with increasing $k$ is shown in Fig. 2(b). The spin-up probability $P_{|\uparrow\downarrow\rangle}$ inhomogeneously oscillates with time. To characterize the inhomogeneity, we transfer Fig. 2(b) to the frequency domain to determine the homogeneity of the exchange oscillation. The frequency spectroscopy is given in Fig. 2(c). When $k$ is approximately $0.06 \sim 0.08 \, \mu s^{-1}$, the spread of the frequency peak is the narrowest, which means that the oscillation homogeneity is best. Thus, we choose $k = 0.07 \, \mu s^{-1}$ in the following experiments, corresponding to a time constant $\tau = \frac{1}{k} \approx 14 \, \mu s$ for the RC circuit. Although the linear ramped top edge relieves the distortion, the exchange oscillation is still disturbed when the oscillation frequency is larger than 20 MHz, as shown in the CPD situation in Fig. 4(a-b). This indicates that the remaining distortion is still influential, and we need a more precise method to calibrate it.

### D. APD method

In the regime of $J \gg \Delta E_z$, the exchange oscillation frequency $f_{ex} \approx J(\varepsilon)$ due to the exponential dependence between $J$ and $\varepsilon$. As $A_{in}$ increases, $\varepsilon$ at the operation point increase as well, and the exchange oscillation frequency increases exponentially. [30]. Therefore, the higher the $A_{in}$ is, the more sensitive the $f_{ex}$ is to pulse distortion. A natural thought is calibrating the control pulse according to the accumulated exchange oscillation phase. However, the phase is susceptible to decoherence and the spin preparation and measurement (SPAM) error. In contrast, the extremum point where the phase is equal to $n*\pi$ ($n = 1,2,3...$) has decoherence and SPAM error insensitive performance and is suitable as a distortion metric. In our experiment, we adjust $A_{in}$ and keep the pulse duration $\lambda$ fixed to obtain the predistorted pulse. When $U_{in}(t)$ is not disturbed, $2\pi f_{ex}(A_{in})\lambda = \pi\ (2\pi)$, and the spin-up probability $P_{|\uparrow\downarrow\rangle}$ should reach extremum points at time $n\lambda$ (n= 1, 2, 3 …). To make the amplitude of the output pulse $U_{out}(t)$ stable during $n\lambda$, we only need to adjust $A_{in}$ in each $\lambda$ and ensure that it corresponds to an extremum point of $P_{|\uparrow\downarrow\rangle}$. Then, we can obtain the output pulse received by the qubits $U_{out}(t)$, which is a precise square wave.

Before correcting the control pulse, we calibrate the relationship between $f_{ex}$ and $A_{in}$ (see Appendix A for more detail), with which we can determine the undistorted $A_{in}$ corresponding to $f_{ex}(A_{in}) = \pi/\lambda$ (or $f_{ex}(A_{in}) = 2\pi/\lambda$). The following steps summarize the experimental procedure:

(1) According to the dependence between $f_{ex}$ and $A_{in}$, we calculate the $A_{in,1}$ that makes the accumulated phase $\phi_{\lambda_1}$ equal to $\pi$ ($2\pi$) at time $\lambda_1$. Here, $\lambda_1$ is the temporal resolution of the calibration procedure, and we assume that the output voltage level during $\lambda_1$ is constant. For example, in the left part of Fig. 3(a), we chose the $A_{in,1}$ corresponding to $f_{ex} = 10$ MHz when ensuring $\lambda_1 = 0.1$ μs to guarantee $\phi_{\lambda_1} = 2\pi$.

(2) Keep the input square pulse duration as $\lambda_1$ and set the pulse amplitude as $c_{11}A_{in,1}$, where $c_{11}$ is the predistortion coefficient. Sweep $c_{11}$ around 1 and find the predistortion coefficient $c_{11}$, which corresponds to the extremum value of $P_{|\uparrow\downarrow\rangle}$. As illustrated in the left part of Fig. 3(a), $c_{11}$ is swept in the interval [0.95, 1.05] and chosen as the value corresponding to the maximum $P_{|\uparrow\downarrow\rangle}$.

(3) Keep the input square pulse $U_{in}(t)$ duration as $2\lambda_1$ and set the pulse amplitude in the time interval $[0, \lambda_1]$ as $c_{11}A_{in,1}$. In the time interval $[\lambda_1, 2\lambda_1]$, we set the pulse amplitude as $c_{12}A_{in,1}$ and sweep $c_{12}$ to approximately 1 to find $c_{12}$ that corresponds to the extremum value of $P_{|\uparrow\downarrow\rangle}$.

(4) Repeat step (3) and determine $c_1 = [c_{11}, c_{12}, c_{13}, ..., c_{1n}]$. The orange pulse segments in the left part of Fig. 3(a) correspond to the calibrated $c_1 = [c_{11}, c_{12}, c_{13}, ..., c_{110}]$.

(5) To increase the calibration temporal resolution, decrease the calibration duration step $\lambda_2$, calculate $A_{in,2}$ and repeat steps (1)-(4) to determine the value of $c_2$. Before this step, set $U_{in}(t)$ as $c_1 u(t)$, and $u(t)$ is the square pulse with amplitude $A_{in,1}$. In the right part of Fig. 3(a), we chose $\lambda_2 = 0.05$ μs, $f_{ex} = 20$ MHz and $\phi_{\lambda_2} = 2\pi$. The calibration of $c_2$ is based on the calibrated pulse in the left part of Fig. 3(a).

(6) Decrease the calibration duration step $\lambda_m$ and calibrate $c_m = [c_{m1}, c_{m2}, c_{m3}, ..., c_{mn}]$ until the temporal resolution is sufficient.

(7) Calculate the predistorted input pulse as $U_{in}(t) = c_1 c_2 ... c_m u(t)$. Finally, we obtain $U_{in}(t)$, which ensures that the pulse delivered into device $U_{out}(t)$ is an exact square pulse.

The APD input pulse $U_{in,APD}(t)$ in our experiment is shown in Fig. 3(b). As the contrast, we also give the CPD pulse $U_{in,CPD}(t)$ with a line ramped top edge and the initial square pulse $U_{in,UPD}(t)$. Two rounds of iterations with $\lambda_1 = 40$ ns and $\lambda_2 = 20$ ns are implemented successively to obtain $U_{in,APD}(t)$. The phase step is $\pi$, and the sweeping interval of $c$ is $[0.85, 1.15]$. During the duration $[0 \text{ ns}, 80 \text{ ns}]$, the pulse amplitude of $U_{in,APD}(t)$ is lower than that of $U_{in,CPD}(t)$ which should correspond to the lower oscillation frequency. In Fig. 4(a), in comparison to the exchange oscillation driven by $U_{in,CPD}(t)$, we observe that the exchange oscillation driven by $U_{in,APD}(t)$ is indeed slower and more homogeneous. Similarly, the pulse amplitude of $U_{in,APD}(t)$ within time [80 ns, 220 ns] is higher, which also improves the oscillation homogeneity compared to $U_{in,\text{CPD}}(t)$. The significant conformity confirms the validity of this procedure. Although the trend is corrected, $U_{in,\text{APD}}(t)$ in Fig. 4(c) is still disturbed by some glitches due to the exchange oscillation instability. To smooth these glitches, we fit $U_{in,\text{APD}}(t)$ with a polynomial curve. We speculate that the glitch problem can be relieved by averaging more time during the calibration or with a higher quality exchange oscillation.

Although the validity is confirmed, this calibration procedure still has some limitations, predominantly in the temporal resolution, longest correction time, and accuracy, and is mainly influenced by the exchange oscillation characteristics. Specifically, the temporal resolution $\lambda_m$ depends on the fastest exchange oscillation frequency, and the coherence time of the exchange oscillation determines the longest correction time $n * \lambda_m$ our procedure can calibrate. The accuracy of this procedure is dominated by the phase step we choose; in other words, the larger the phase step we choose, the more susceptible the extremum point is. However, for the same $\lambda_m$, a larger phase step means that a higher $f_{ex}$ is needed, which is often accompanied by a faster decoherence speed and results in a shorter correction time. Thus, these parameters should be chosen carefully.

Remarkably, two situations could lead to the failure of the APD procedure. One is the multiple peaks in the scanning interval of $c$. As shown in Fig. 3(c), while modifying $c_{mn}$, we observe that the oscillation frequency increases gradually. For the different pulse durations $\lambda$ in Fig. 3(c), the latest extremum points exist when $A_{in0}$ shifts $0.48 A_{in0}$, $0.18 A_{in0}$ and $0.14 A_{in0}$, respectively. $A_{in0}$ is the ideal pulse amplitude corresponding to the extremum point. A higher temporal resolution as well as a smaller $\lambda$ are more likely to lead to multiple peaks. Once the sweep interval is too large, multiple peaks will appear and result in a mistakable judgment of $c_{mn}$. Another situation can lead to the APD procedure failing when the distortion is so strong that we detect the wrong probability peak. Paradoxically, a narrow sweeping interval of $c_{mn}$ is encouraged to avoid the multipeak phenomenon, but it can make us miss the extremum point when the distortion is strong. The solution that we provide is starting with a longer temporal resolution as well as a larger λ, which is commonly companied by a smaller average distortion and less disturbance by the multipeak phenomenon, so that we can choose a narrower sweeping interval and then increase the resolution during multiple iterations. Or use the CPD procedure before the APD procedure. In general, the temporal resolution, $\lambda_m$, $n$, and c should be carefully selected to ensure the validity of the calibration procedure according to the exchange oscillation characteristics.

### E. Validity of predistortion methods

With $U_{in,\text{APD}}(t)$, we are able to easily calculate the transfer function $h(t)$ and the inverse transfer function $h^{-1}(t) = U'_{in,\text{APD}}(t)$ and obtain the predistorted pulse for an arbitrary output pulse as $U_{in,\text{APD}}(t) = h^{-1}(t) * U_{out}(t)$ (see Appendix B for more detail). In Fig. 4(a), we demonstrate the time dependence of the exchange oscillation with $U_{in,\text{UPD}}(t)$, $U_{in,\text{CPD}}$ and $U_{in,\text{APD}}$ as a function of $A_{in}$. To indicate the homogeneity of the oscillation, we transfer the exchange oscillation from the time domain to the frequency domain, as shown in Fig. 4(b). When the UPD pulse $U_{in,\text{UPD}}(t)$ is exerted on the qubits, as $A_{in}$ increases, the frequency spectrum of the exchange oscillation becomes vague. And as we mentioned before, $f_{ex}$ gradually slows down with time, mainly affected by the RC filter circuit. Hence, we correct the distortion with a linear ramp top edge. As shown in Fig. 4(b), in the situation of CPD, the frequency spectroscopy is improved when $f_{ex}$ is lower than 20 MHz, while it is still vague at a higher frequency regime. This indicates that the distortion is partly relieved while the influence of the residual part is still noticeable. With the APD procedure, the frequency spectroscopy is clear even when $f_{ex}$ is higher than 20 MHz, and the homogeneity is significantly improved. The residual width of the frequency peak is due to oscillation decoherence and low temporal resolution, which is 50 MHz here.

The calibrated pulse supports the characterization of the coherence time and influences further research on the error source. In contrast to the CPD situation, the APD procedure calibrates the distortion in the first several period oscillations, which is

9/19point when the distortion is strong. The solution that we provide is starting with a longer temporal resolution as well as a larger λ, which is commonly companied by a smaller average distortion and less disturbance by the multipeak phenomenon, so that we can choose a narrower sweeping interval and then increase the resolution during multiple iterations. Or use the CPD procedure before the APD procedure. In general, the temporal resolution, $\lambda_m$, $n$, and c should be carefully selected to ensure the validity of the calibration procedure according to the exchange oscillation characteristics.

### E. Validity of predistortion methods

With $U_{in,\text{APD}}(t)$, we are able to easily calculate the transfer function $h(t)$ and the inverse transfer function $h^{-1}(t) = U'_{in,\text{APD}}(t)$ and obtain the predistorted pulse for an arbitrary output pulse as $U_{in,\text{APD}}(t) = h^{-1}(t) * U_{out}(t)$ (see Appendix B for more detail). In Fig. 4(a), we demonstrate the time dependence of the exchange oscillation with $U_{in,\text{UPD}}(t)$, $U_{in,\text{CPD}}$ and $U_{in,\text{APD}}$ as a function of $A_{in}$. To indicate the homogeneity of the oscillation, we transfer the exchange oscillation from the time domain to the frequency domain, as shown in Fig. 4(b). When the UPD pulse $U_{in,\text{UPD}}(t)$ is exerted on the qubits, as $A_{in}$ increases, the frequency spectrum of the exchange oscillation becomes vague. And as we mentioned before, $f_{ex}$ gradually slows down with time, mainly affected by the RC filter circuit. Hence, we correct the distortion with a linear ramp top edge. As shown in Fig. 4(b), in the situation of CPD, the frequency spectroscopy is improved when $f_{ex}$ is lower than 20 MHz, while it is still vague at a higher frequency regime. This indicates that the distortion is partly relieved while the influence of the residual part is still noticeable. With the APD procedure, the frequency spectroscopy is clear even when $f_{ex}$ is higher than 20 MHz, and the homogeneity is significantly improved. The residual width of the frequency peak is due to oscillation decoherence and low temporal resolution, which is 50 MHz here.

The calibrated pulse supports the characterization of the coherence time and influences further research on the error source. In contrast to the CPD situation, the APD procedure calibrates the distortion in the first several period oscillations, which is crucial for improving the gate fidelity. However, considering the complexity and calibration time, whether CPD or APD procedures are appropriate depends on the experimental requirements. The CPD procedure is faster while compromising in accuracy, while the APD procedure is more precise, which is sustained by multi-round iteration. When $f_{ex} < 20$ MHz, the calibration accuracy of the CPD method and the APD method are compactable, and the frequency peak width deviation can come from the occasional un-averaged disturbed oscillation.

## III. CONCLUSION

In summary, using the two-qubit exchange oscillation in a silicon metal–oxide–semiconductor (MOS) quantum dot as the detector, we demonstrate two different methods to calibrate the distortion in our system. The CPD procedure with a linear ramp top edge quickly calibrates the distortion from the bias-tee. The APD procedure, which uses the extremum value of $P_{|\uparrow\downarrow\rangle}$ as the metric, characterizes the transfer function of the measurement system. We observe a significant improvement in the exchange oscillation homogeneity with the transfer function in all frequency ranges. These methods are straightforward and appropriate for any qubit system with oscillation driven by the diabatic square pulse. We believe this correction procedure can potentially improve the gate fidelity in future experiments.

## ACKNOWLEDGMENTS

This work was supported by the National Natural Science Foundation of China (Grants No. 12074368, 92165207, 12034018 and 92265113), the Innovation Program for Quantum Science and Technology (Grant No. 2021ZD0302300), the Anhui Province Natural Science Foundation (Grants No. 2108085J03), and the USTC Tang Scholarship. This work was partially carried out at the USTC Center for Micro and Nanoscale Research and Fabrication.

## APPENDIX A: Calibrating the relationship between $f_{ex}$ and $A_{in}$

Benefiting from the waveform distortion sensitivity, the extremum points of $P_{|\uparrow\downarrow\rangle}$ during the exchange oscillation are used as the metric in our calibration procedure. Before the calibration, a necessary precondition is knowing the $A_{in}$ corresponding to the needed $f_{ex}$. In other words, we need to obtain the relationship between $A_{in}$ and $f_{ex}$ before calibration. However, the distorted pulse disturbs the exchange oscillation, making the directly detected relationship between $A_{in}$ and $f_{ex}$ inaccurate. Although $f_{ex}$ equals the energy splitting $\Delta E$ between the antiparallel states $|\uparrow\downarrow\rangle$ and $|\downarrow\uparrow\rangle$, and $\Delta E$ can be detected accurately, to generate the control pulses faster, the waveform that detects $\Delta E$ and drives the exchange oscillation is generated by different AWGs with different clock rates as well as delivered from the different transfer lines, and we still cannot directly measure the relationship between $A_{in}$ and $f_{ex}$. To address this problem, in our experiment, we determine the dependence of $A_{in,\text{AWG1}}$ on $f_{ex}$ and the conversion relationship between $A_{in,\text{AWG1}}$ and $A_{in,\text{AWG2}}$ such that we obtain $A_{in,\text{AWG2}}(f_{ex})$ indirectly. Here, $A_{in,\text{AWG1}}$ and $A_{in,\text{AWG2}}$ express the amplitudes of the pulse generated from AWG1 and AWG2, respectively.

In Fig. 5(a), we extract $\Delta E$ as increases $A_{in,\text{AWG1}}$ and then fit $A_{in,\text{AWG1}}$ as a function of $\Delta E$, which is equal to $f_{ex}$. $A_{in,\text{AWG1}}$ increases logarithmically with $f_{ex}$, which agrees with the exponential dependence between $J$ and $\varepsilon$. We measure the time dependence of the exchange oscillation while increasing $A_{in,\text{AWG2}}$. $f_{ex}$ can be fitted as a function of $A_{in,\text{AWG2}}$, as shown in Fig. 5(b). It is notable that the fitted $f_{ex}$ is inaccurate because of the disturbed waveform. With the relationship between $A_{in,\text{AWG1}}$ and $f_{ex}$, we can calculate $A_{in,\text{AWG1}}$, which corresponds to $f_{ex}$, and then fit the dependence between $A_{in,\text{AWG1}}$ and $A_{in,\text{AWG2}}$ with a linear function, as given in Fig. 5(c). With the two fitted functions, the deviation origin from the distortion is made up, and $A_{in,\text{AWG2}}$, which is $A_{in}$ in the main article, corresponding to arbitrary $f_{ex}$ is accurately determined.

## APPENDIX B: Calculating the transfer function

In class control theory, for a linear time-invariant system, the response signal can be determined by the transform function to an arbitrary input signal. We assume that the circuit delivers the control pulse as a linear time-invariant system. Hence, the response signal $U_{out}(t)$ to an arbitrary input signal $U_{in}(t)$ can be determined by the transform function $h(t)$ with the following relationship:

$$U_{out}(t) = \int_0^\infty h(t-\tau)U_{in}(t)d\tau = h(t) * U_{in}(t). \qquad (1)$$

Actually, we want to determine the predistorted input waveform $U_{in}(t)$ to an arbitrary output waveform. In other words, we want to find the inverse transfer function $h^{-1}(t)$ such that:

$$U_{in}(t) = \int_0^\infty h^{-1}(t-\tau)U_{out}(t)d\tau = h^{-1}(t) * U_{out}(t). \qquad (2)$$

The inverse transfer function $h^{-1}(t)$ can be measured by choosing the output signal $U_{out}(t)$ as a δ-function, from which we can obtain $U_{in}(t) = h^{-1}(t)$. However, characterizing $U_{out}(t)$ as a δ-function is impossible to realize. And as alternatives, we choose $U_{out}(t)$ as a step function, and the inverse transfer function is given as $h^{-1}(t) = U'_{in}(t)$. With our predistortion procedure, the calculated $U_{in}(t)$ corresponds to $U_{out}(t)$, which is a precise square function. Therefore, after fitting $U_{in}(t)$ with a polynomial function, we fix the top edge of $U_{out}(t)$ in $t = 0$, and the inverse transfer function can be obtained as $U'_{in}(t)$.

**Figure Captions:**

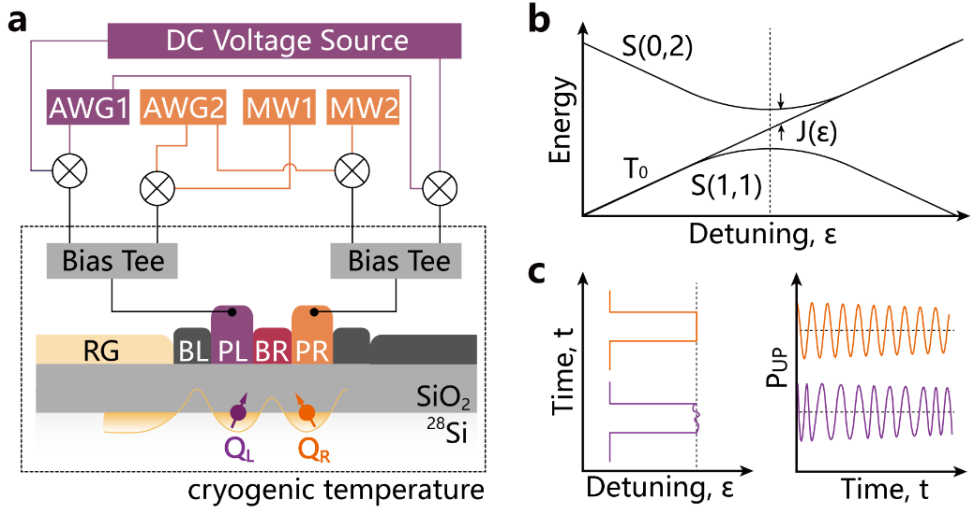

**FIG. 1.** (a) Schematic of the measurement system and device. The pulse on the order of kHz is generated by AWG1 overlaps with the DC voltage. And the pulse on the order of GHz is generated by AWG2 and overlaps with the microwave. At cryogenic temperature, two bias tees combine all control signals and deliver them to the plunger gates PL and PR, respectively. The two qubits $Q_L$ and $Q_L$ are located underneath the two plunger gates and are directly controlled by the signal output from the bias tees. (b) Energy-level diagram of the two-qubit system as a function of the detuning $\varepsilon$. $\varepsilon$ is controlled by the relative voltage between PL and PR. The exchange coupling $J(\varepsilon)$ is the energy splitting between the singlet and triplet states. (c) Illustration of the exchange oscillation corresponding to the undistorted (orange) and distorted (purple) pulses. The exchange oscillation frequency $f_{ex}$ is proportional to the pulse amplitude at time *t*.

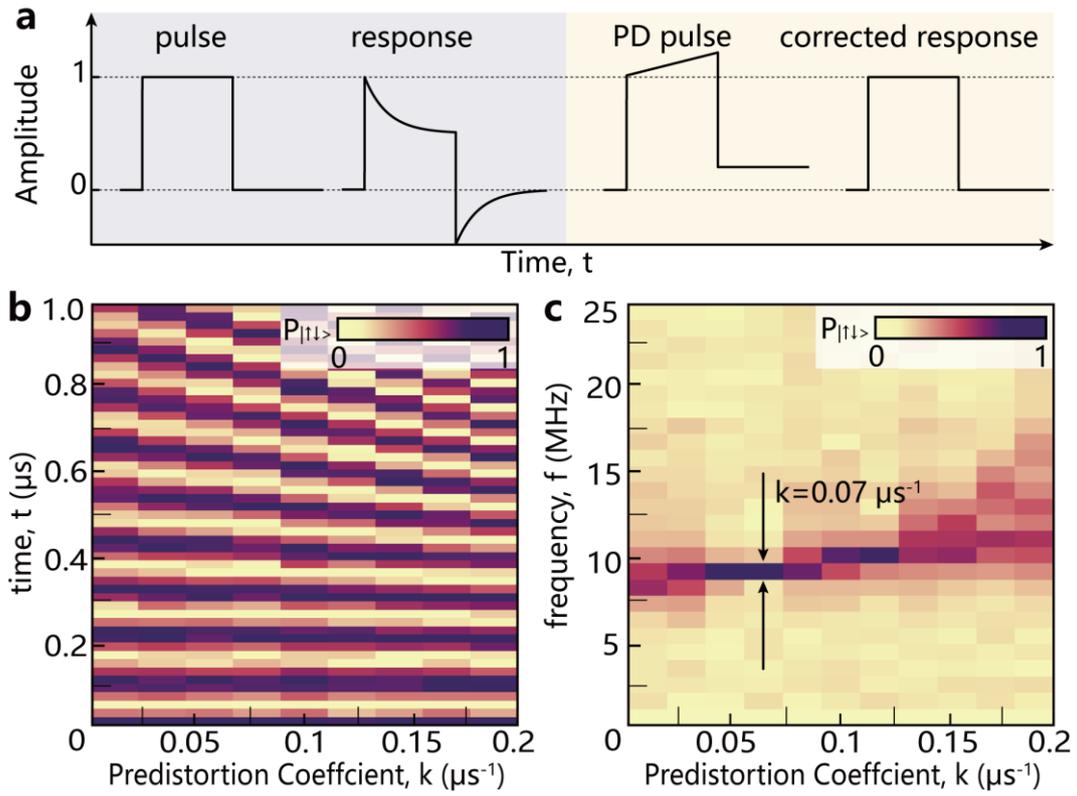

**FIG. 2.** (a) Schematic of the square pulse response after the RC filter circuit inside the bias tee. The predistorted (PD) pulse with a linearly ramped top edge can correct the distortion. (b) Time-dependent responses of the spin-up probability of the $|\uparrow\downarrow\rangle$ state with the predistorted pulse. The oscillation frequency significantly increases as the predistortion coefficient $k$ increases. The frequency spectroscopy in (c) gives the Fourier transformation of (b). The narrow FHWM around $k = 0.07\ \mu s^{-1}$ indicates improved oscillation homogeneity.

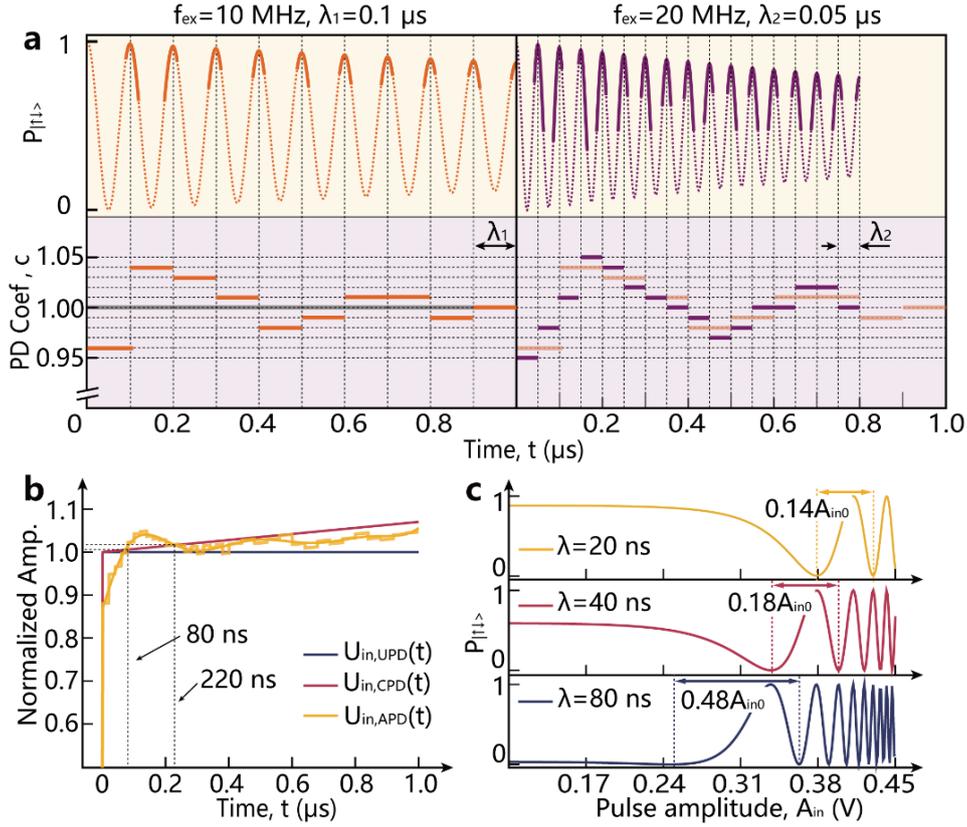

**FIG. 3.** (a) Schematic of our calibration procedure. The spin-up probability $P_{|\uparrow\downarrow\rangle}$ is measured as a function of the predistortion coefficient c. During each time interval $\lambda_n$, $c_n$ is chosen to maximize $P_{|\uparrow\downarrow\rangle}$, and the amplitude of $U_{in}(t)$ is adjusted to $c_n * A_{in}$. Here are two rounds of calibration that chose the phase step $\phi_\lambda$ as $2\pi$ and the sweeping interval of the pulse amplitude as $\pm 15\%$. According to the characteristics of exchange oscillation, we apply $\lambda_1 = 0.1$ μs, $n_1 = 10, \lambda_2 = 0.05$ μs, and $n_2 = 16$. (b) The initial square pulse $U_{in,\text{UPD}}(t)$ (blue), the coarse predistorted pulse (CPD) $U_{in,\text{CPD}}(t)$ (red) and the all predistorted input pulse $U_{in,\text{APD}}(t)$ (yellow). The linearly ramped rate k of $U_{in,\text{CPD}}(t)$ is 0.07 μs$^{-1}$. Two rounds of iterations with $\lambda_1 = 40$ ns and $\lambda_2 = 20$ ns are implemented to calibrate $U_{in,\text{APD}}(t)$. The polynomial curve fits $U_{in,\text{APD}}(t)$ to relieve the glitches. The dashed lines indicate the time at 80 ns and 220 ns, correspond to where $U_{in,\text{CPD}}(t) = U_{in,\text{APD}}(t)$. All three pulses are normalized. (c) The multipeak phenomenon during calibration. As $A_{in}$ is modified, the smaller temporal resolution λ corresponding to a larger $f_{ex}$ is more susceptible to the multipeak phenomenon. The colored dashed lines indicate $A_{in}$, where the accumulated phase equals π and 3π, respectively.

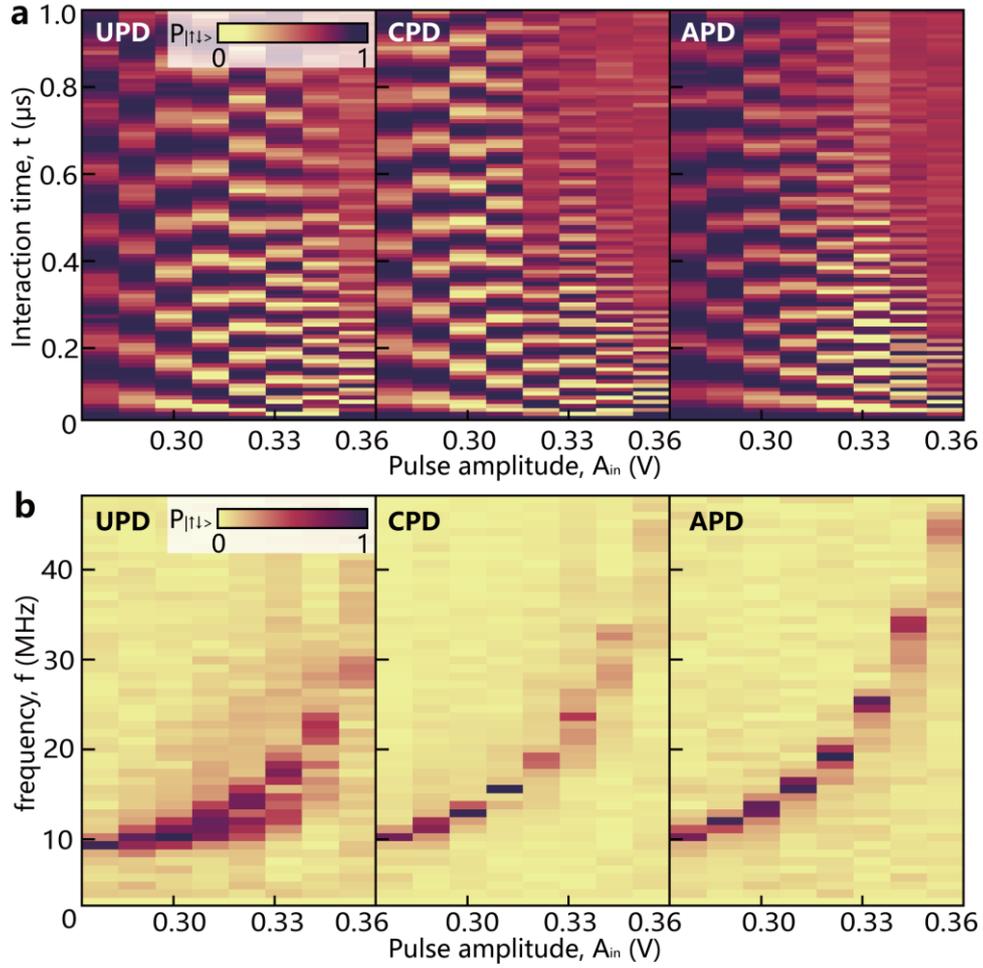

**FIG. 4.** (a) The time dependence of the exchange oscillation between the $|\uparrow\downarrow\rangle$ and $|\downarrow\uparrow\rangle$ states of two spin qubits driven with the UPD pulse $U_{in,\text{UPD}}(t)$, the CPD pulse $U_{in,\text{CPD}}(t)$ and the APD pulse $U_{in,\text{APD}}(t)$. (b) Frequency spectroscopy is obtained by transferring the exchange oscillation in (a) into the frequency domain.

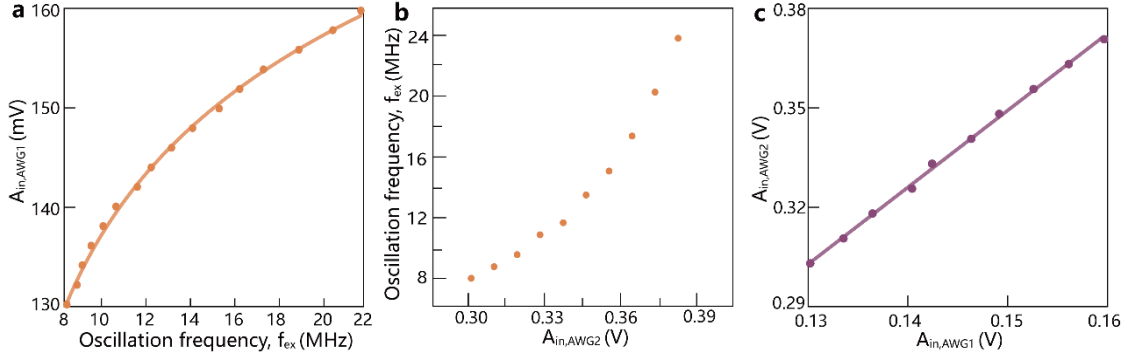

**FIG. 5.** (a) The relationship between the exchange oscillation frequency $f_{ex}$ and the output amplitude of AWG1 $A_{in,\text{AWG1}}$. The experimental data are fitted with a logarithmic function. (b) $f_{ex}$ as a function of the output amplitude of AWG2 $A_{in,\text{AWG2}}$. $f_{ex}$ is extracted by fitting the oscillation with $P_{|\uparrow\downarrow\rangle} = A * e^{-\frac{t}{T_{2,ex}}} * \cos(2\pi f_{ex} t + \varphi) + y_0$. $T_{2,ex}$ is the coherent time of the exchange oscillation. With the function fitted in (a), the $A_{in,\text{AWG1}}$ corresponding to $f_{ex}$ is obtained. Combined with the dependence between $f_{ex}$ and $A_{in,\text{AWG2}}$ in (b), we obtain the linear relationship between $A_{in,\text{AWG1}}$ and $A_{in,\text{AWG2}}$ in (c).

# References

[1] D.P. DiVincenzo, The Physical Implementation of Quantum Computation, Fortschritte der Physik, **48**, 771 (2000).

[2] G. Burkard, T.D. Ladd, A. Pan, J.M. Nichol, J.R. Petta, Semiconductor spin qubits, Rev. Mod. Phys. **95**, 025003 (2023).

[3] X. Zhang, H.-O. Li, G. Cao, M. Xiao, G.-C. Guo, G.-P. Guo, Semiconductor quantum computation, National Science Review **6**, 32 (2019).

[4] F.A. Zwanenburg, A.S. Dzurak, A. Morello, M.Y. Simmons, L.C.L. Hollenberg, G. Klimeck, S. Rogge, S.N. Coppersmith, M.A. Eriksson, Silicon quantum electronics, Rev. Mod. Phys. **85**, 961 (2013).

[5] M. Veldhorst, C.H. Yang, J.C.C. Hwang, W. Huang, J.P. Dehollain, J.T. Muhonen, S. Simmons, A. Laucht, F.E. Hudson, K.M. Itoh, A. Morello, A.S. Dzurak, A two-qubit logic gate in silicon, Nature **526**, 410 (2015).

[6] T.F. Watson, S.G.J. Philips, E. Kawakami, D.R. Ward, P. Scarlino, M. Veldhorst, D.E. Savage, M.G. Lagally, M. Friesen, S.N. Coppersmith, M.A. Eriksson, L.M.K. Vandersypen, A programmable two-qubit quantum processor in silicon, Nature **555**, 633 (2018).

[7] D.M. Zajac, A.J. Sigillito, M. Russ, F. Borjans, J.M. Taylor, G. Burkard, J.R. Petta, Resonantly driven CNOT gate for electron spins, Science **359**, 439 (2018).

[8] X. Xue, M. Russ, N. Samkharadze, B. Undseth, A. Sammak, G. Scappucci, L.M.K. Vandersypen, Quantum logic with spin qubits crossing the surface code threshold, Nature **601**, 343 (2022).

[9] A. Noiri, K. Takeda, T. Nakajima, T. Kobayashi, A. Sammak, G. Scappucci, S. Tarucha, Fast universal quantum gate above the fault-tolerance threshold in silicon, Nature **601**, 338 (2022).

[10] A.R. Mills, C.R. Guinn, M.J. Gullans, A.J. Sigillito, M.M. Feldman, E. Nielsen, J.R. Petta, Two-qubit silicon quantum processor with operation fidelity exceeding 99%, Sci. Adv. **8**, eabn5130 (2022).

[11] C.H. Yang, K.W. Chan, R. Harper, W. Huang, T. Evans, J.C.C. Hwang, B. Hensen, A. Laucht, T. Tanttu, F.E. Hudson, S.T. Flammia, K.M. Itoh, A. Morello, S.D. Bartlett, A.S. Dzurak, Silicon qubit fidelities approaching incoherent noise limits via pulse engineering, Nature Electronics **2**, 151 (2019).

[12] K. Takeda, J. Yoneda, T. Otsuka, T. Nakajima, M.R. Delbecq, G. Allison, Y. Hoshi, N. Usami, K.M. Itoh, S. Oda, T. Kodera, S. Tarucha, Optimized electrical control of a Si/SiGe spin qubit in the presence of an induced frequency shift, Npj Quantum Inform. **4**, 54 (2018).

[13] T. Struck, A. Hollmann, F. Schauer, O. Fedorets, A. Schmidbauer, K. Sawano, H. Riemann, N.V. Abrosimov, Ł. Cywiński, D. Bougeard, L.R. Schreiber, Low-frequency spin qubit energy splitting noise in highly purified 28Si/SiGe, Npj Quantum Inform. **6**, 40 (2020).

[14] O.E. Dial, M.D. Shulman, S.P. Harvey, H. Bluhm, V. Umansky, A. Yacoby, Charge Noise Spectroscopy Using Coherent Exchange Oscillations in a Singlet-Triplet Qubit, Phys. Rev. Lett. **110**, 146804 (2013).

[15] J. Yoneda, K. Takeda, T. Otsuka, T. Nakajima, M.R. Delbecq, G. Allison, T. Honda, T. Kodera, S. Oda, Y. Hoshi, N. Usami, K.M. Itoh, S. Tarucha, A quantum-dot spin qubit with coherence limited by charge noise and fidelity higher than 99.9%, Nat. Nanotechnol. **13**, 102 (2018).

[16] E. Kawakami, T. Jullien, P. Scarlino, D.R. Ward, D.E. Savage, M.G. Lagally, V.V. Dobrovitski, M. Friesen, S.N. Coppersmith, M.A. Eriksson, L.M.K. Vandersypen, Gate fidelity and coherence of an electron spin in an Si/SiGe quantum dot with micromagnet, Proceedings of the National Academy of Sciences **113**, 11738 (2016).

[17] J. Bylander, M.S. Rudner, A.V. Shytov, S.O. Valenzuela, D.M. Berns, K.K. Berggren, L.S. Levitov, W.D. Oliver, Pulse imaging and nonadiabatic control of solid-state artificial atoms, Phys. Rev. B **80**, 220506 (2009).

[18] S. Gustavsson, O. Zwier, J. Bylander, F. Yan, F. Yoshihara, Y. Nakamura, T.P. Orlando, W.D. Oliver, Improving Quantum Gate Fidelities by Using a Qubit to Measure Microwave Pulse Distortions, Phys. Rev. Lett. **110**, 040502 (2013).

[19] M. Jerger, A. Kulikov, Z. Vasselin, A. Fedorov, In Situ Characterization of Qubit Control Lines: A Qubit as a Vector Network Analyzer, Phys. Rev. Lett. **123**, 150501 (2019).

[20] M.A. Rol, L. Ciorciaro, F.K. Malinowski, B.M. Tarasinski, R.E. Sagastizabal, C.C. Bultink, Y. Salathe, N. Haandbaek, J. Sedivy, L. DiCarlo, Time-domain characterization and correction of on-chip distortion of control pulses in a quantum processor, Appl. Phys. Lett. **116**, 054001 (2020).

[21] N.K. Langford, R. Sagastizabal, M. Kounalakis, C. Dickel, A. Bruno, F. Luthi, D.J. Thoen, A. Endo, L. DiCarlo, Experimentally simulating the dynamics of quantum light and matter at deep-strong coupling, Nature Communications **8**, 1715 (2017).

[22] B.R. Johnson, Controlling Photons in Superconducting Electrical Circuits, Yale University (2011).

[23] J. Kelly, Fault-tolerant superconducting qubits, University of California, Santa Barbara (2015).

[24] R.-Z. Hu, R.-L. Ma, M. Ni, Y. Zhou, N. Chu, W.-Z. Liao, Z.-Z. Kong, G. Cao, G.-L. Wang, H.-O. Li, G.-P. Guo, Flopping-mode spin qubit in a Si-MOS quantum dot, Appl. Phys. Lett. **122**, 134002 (2023).

[25] X. Zhang, Y. Zhou, R.-Z. Hu, R.-L. Ma, M. Ni, K. Wang, G. Luo, G. Cao, G.-L. Wang, P. Huang, X. Hu, H.-W. Jiang, H.-O. Li, G.-C. Guo, G.-P. Guo, Controlling Synthetic Spin-Orbit Coupling in a Silicon Quantum Dot with Magnetic Field, Phys. Rev. Appl. **15**, 044042 (2021).

[26] X. Zhang, R.-Z. Hu, H.-O. Li, F.-M. Jing, Y. Zhou, R.-L. Ma, M. Ni, G. Luo, G. Cao, G.-L. Wang, X. Hu, H.-W. Jiang, G.-C. Guo, G.-P. Guo, Giant Anisotropy of Spin Relaxation and Spin-Valley Mixing in a Silicon Quantum Dot, Phys. Rev. Lett. **124**, 257701 (2020).

[27] J. Yoneda, T. Otsuka, T. Takakura, M. Pioro-Ladrière, R. Brunner, H. Lu, T. Nakajima, T. Obata, A. Noiri, C.J. Palmstrøm, A.C. Gossard, S. Tarucha, Robust micromagnet design for fast electrical manipulations of single spins in quantum dots, Appl. Phys. Express **8**, 084401 (2015).

[28] M. Russ, D.M. Zajac, A.J. Sigillito, F. Borjans, J.M. Taylor, J.R. Petta, G. Burkard, High-fidelity quantum gates in Si/SiGe double quantum dots, Phys. Rev. B **97** (2018).

[29] J. R. Petta, A. C. Johnson, J. M. Taylor, E. A. Laird, A. Yacoby, M. D. Lukin, C. M. Marcus, M. P. Hanson, and A. C. Gossard, Coherent Manipulation of Coupled Electron Spins in Semiconductor Quantum Dots, Science **309**, 2180 (2005).

[30] T.F. Watson, S.G.J. Philips, E. Kawakami, D.R. Ward, P. Scarlino, M. Veldhorst, D.E. Savage, M.G. Lagally, M. Friesen, S.N. Coppersmith, M.A. Eriksson, L.M.K. Vandersypen, A programmable two-qubit quantum processor in silicon, Nature **555**, 633 (2018).